\documentclass[twocolumn,showpacs,floatfix,aps]{revtex4}
\usepackage[hypertex]{hyperref}
\usepackage{graphicx}
\usepackage{dcolumn}
\usepackage{bm}

\def\deltaf{\delta_{\rm f}}
\def\ef{{\bf E}_{\rm f}}

\begin{document}

\title{Low-frequency conductivity of a non-degenerate
  2D electron liquid\\ in strong magnetic fields}

\author{M.I.~Dykman}
\email[]{dykman@pa.msu.edu}
\affiliation{Department of Physics and Astronomy and Institute
  for Quantum Sciences, Michigan State University, East Lansing, MI
  48824}

\author{Leonid P.~Pryadko}
\email[]{leonid@landau.ucr.edu}

\affiliation{Department of Physics, University of California,
 Riverside, CA 92521} 
 \date\today
\begin{abstract}
  We study the conductivity of a nondegenerate 2D electron liquid in a
  quantizing magnetic field for frequencies well below the cyclotron
  frequency.  The conductivity is formed by electron transitions in
  which the energy of a photon goes to the interaction energy of the
  many-electron system, whereas the involved momentum is transferred
  to quenched disorder. The conductivity peak is non-Lorentzian. Its
  shape depends on the relation between the correlation length $r_{\rm
    c}$ of the disorder potential and the typical amplitude
  $\delta_{\rm f}$ of vibrations of the electrons about their
  quasi-equilibrium positions in the liquid.  The width of the peak is
  determined by the reciprocal time it takes an electron to move over
  $r_{\rm c}$ (or the magnetic length $l$, for $r_{\rm c}< l$). In
  turn, this time is determined by vibrational or diffusive motion,
  depending on the ratio $r_c/\delta_{\rm f}$.  We analyze the tail of the
  conductivity peak for short-range disorder. It is formed by multiple
  collisions with the disorder potential.  We also analyze scattering
  by rare negatively charged traps and show that the conductivity
  spectrum in this case depends on both short- and long-time electron
  dynamics.
\end{abstract}

\maketitle

\section{Introduction}

In recent years much progress has been made toward understanding of
transport phenomena in strongly interacting electron systems. The
well-known examples are the fractional quantum Hall effect (QHE)
\cite{FQHE} and metal-insulator transition phenomena in low-density
two-dimensional electron systems (2DES) in semiconductors
\cite{Abrahams-00}.  2DESs are particularly convenient for
investigating the electron-electron interaction (EEI), since the
electron density $n$ can be varied in broad limits. One of the most
important effects of the EEI is onset of electron correlations. The
extent to which the system is correlated depends on the ratio $\Gamma$
of the characteristic Coulomb energy $e^2(\pi n)^{1/2}$ to the
electron kinetic energy $E_{\rm kin}$
\begin{equation}
\label{Gamma}
\Gamma = e^2(\pi n)^{1/2}/E_{\rm kin}
\end{equation}
($E_{\rm kin}$ is equal to the biggest of the Fermi energy
$\epsilon_F$ and $k_BT$). For $\Gamma$ exceeding a critical
value $\Gamma_W$ a 2DES becomes a Wigner crystal.  The parameter
$\Gamma_W$ is numerically large. For low temperatures ($E_{\rm
kin}=\epsilon_F\gg k_BT$, in which case $\Gamma = r_s$),
$\Gamma_W\approx 37$ \cite{Tanatar-89}, whereas for a nondegenerate
2DES ($\epsilon_F\ll k_BT$) $\Gamma_W\approx 130$ \cite{Andrei-Book}.

For $\Gamma_W> \Gamma\gg 1$, a 2DES is still strongly correlated, but
it forms an electron liquid. Based on the success of the Fermi liquid
theory in describing $^3$He it is often assumed that, in the quantum
region $\epsilon_F\gg k_BT$, the electron liquid can be described by a
Fermi liquid, too.
However, for
very large $r_s$, the system may be more complicated \cite{Tanatar-89,ferro}.

For large $k_BT/\epsilon_F$ and for $\Gamma_W> \Gamma\gg 1$, a 2DES is
a strongly correlated non-Fermi liquid. It should display a
nonstandard behavior. Experimentally such 2DES has been investigated
in semiconductor heterostructures \cite{Abrahams-00,hetero-ND}, and in
much detail for electrons on the surface of liquid helium
\cite{Andrei-Book}. A nondegenerate electron liquid has common
features with an ordinary liquid. It does not have long-range
translational order and displays self-diffusion, as seen in various
numerical simulations
\cite{Hansen79,Kalia81,Strandburg88,Fang-Yen-97}.

In contrast to ordinary liquids driven by forces applied to their
surface, a 2DES is driven by a ``volume'' field, that is a field
experienced by each electron. It comes from external disorder, such as
defects and phonons in semiconductors, or helium vapor atoms and
surface capillary waves on helium. Because the total momentum of a 2D
electron liquid is changed through ``volume'' rather than ``surface''
scattering, electron transport is different from transport in an
ordinary liquid. It is different also from transport in a weakly
nonideal electron gas. Even though the conductivity is of a metallic
rather than activated type for weak disorder potential, its dependence
on the parameters of the system, including electron density,
temperature, frequency, and a magnetic field, should be totally
different from that for an ideal gas.

The change of transport coefficients stems from electron scattering by
disorder being strongly affected by the electron-electron interaction
in a 2D liquid.  During scattering an electron is coupled to other
electrons, and this coupling determines the scattering
probability. This is the case even for a short-range disorder
potential, where different electrons are scattered off uncorrelated
fluctuations. In this respect a nondegenerate electron liquid is
similar to a liquid of vortices in a superconductor.

In the present paper we investigate the frequency dependence of the
magneto-conductivity
$\sigma_{xx}(\omega)$ of a 2D electron liquid. Such dependence is
particularly interesting as it should provide a direct insight into
the way in which a correlated electron system exchanges momentum with
disorder. Indeed, in the Drude model the low-frequency
magneto-conductivity transverse to a strong magnetic field $B$ is
\begin{equation}
\label{eq:drude}
\sigma_{\rm Dr}(\omega)={ne^2\tau^{-1}_{\rm Dr}(\omega) /
m\omega_c^2},\quad \omega_c=eB/mc,
\end{equation}
where $\omega_c$ is the cyclotron frequency and $\tau^{-1}_{\rm
Dr}$ is the momentum relaxation rate. Here and below we keep in
$\sigma_{xx}(\omega)$ the lowest-order term in $\omega_c^{-1}$. The
frequency dispersion of $\sigma_{\rm Dr}$ comes from the dispersion of
$\tau^{-1}_{\rm Dr}(\omega)$. In the case of
single-electron elastic scattering it becomes strong for frequencies
of order of the duration of a collision with a scatterer.

The single-electron Drude picture does not apply to a 2DES if the
magnetic field $B$ transverse to the electron layer is
quantizing. Here, all single-electron states except for one or maybe
few are localized \cite{Pruisken-localization,Huckestein-95}, and the
single-electron conductivity is equal to zero at zero
frequency\cite{Wang-Fisher-Girvin-Chalker,Kuehnel01}.  Metallic-like
conductivity of a nondegenerate electron liquid is a result of the
electron-electron interaction. Earlier we found a way to take this
interaction into account nonperturbatively and calculated the static
conductivity and the cyclotron resonance for a weak short-range
disorder potential \cite{Dykman-All}. These results were qualitatively
and quantitatively confirmed by the experiment \cite{Lea-97,Teske99}.

The question of observing the actual dynamics or electron scattering
in the electron liquid has not been addressed previously. From analogy
with single-electron scattering one may expect that an insight into
this dynamics can be gained from the frequency dependence of
$\sigma_{xx}(\omega)$. Here we develop an appropriate theory and
suggest relevant experiments. We also study the magnetoconductivity
for two important types of a disorder potential that have not been
discussed for a nondegenerate electron liquid: a smooth
random potential and a potential of rare charged defects. The
physics of many-electron transport in these cases is significantly
different from that for short-range disorder.

A simple argument shows that the frequency dispersion of the
long-wavelength magneto-conductivity for $\omega\ll \omega_c$ is
indeed directly related to many-electron effects.  Because the
electron kinetic energy is quantized, the energy of an absorbed photon
$\hbar\omega$ may go to either the electron potential energy in the
disorder potential or the Coulomb energy of the electron system, or
both. For weak disorder electrons are not localized, and the disorder
potential is largely averaged out by electron motion. Then the photon
energy may only be transferred to the electron system. However, to
provide momentum conservation, this transfer must be mediated by
disorder.

For short-range disorder, one can think of photon absorption as
resulting from an electron bouncing off a point defect.  In a quantizing
magnetic field ${\bf B}$, a momentum transfer to a defect ${\delta \bf
p}$ leads to an electron displacement $\delta {\bf r}=
(c/eB^2)\,{\delta \bf p\times B}$. In the presence of radiation, this
displacement can change the energy of the electron-electron
interaction by $\hbar\omega$.  Therefore by investigating the
frequency dependence of the absorption cross-section, one can find how
an electron moves during a collision.

As we show, for weak short-range disorder the conductivity
$\sigma_{xx}(\omega)$ has a peak at $\omega=0$ with a specific
non-Lorentzian shape, which is fully determined by the
electron-electron interaction. Of special interest is the tail of this
peak.  For large photon frequencies (but still $\omega \ll \omega_c$)
a large electron displacement $\delta {\bf r}$ is required in order to
accommodate the energy. Respectively, a large momentum has to be
transferred to the disorder. It may come from multiple electron
scattering.  The mechanism has some similarity with that of anomalous
diffusion transverse to a magnetic field \cite{Thouless}. It
significantly slows down the decay of $\sigma_{xx}(\omega)$ with
$\omega$ compared to the decay calculated for single-collision
absorption.

The frequency dependence of $\sigma_{xx}(\omega)$ is totally different
for a long-range disorder potential. Of particular interest is the
potential with a correlation length $r_{\rm c}$ smaller than the
inter-electron distance. In this case the electron scattering cannot
be described in the hydrodynamic approximation. As we show, here the
conductivity displays a characteristic cusp at $\omega=0$.

In Sec.~II we relate the magnetoconductivity of a strongly correlated
electron liquid to the electron structure factor. We introduce the
single-site approximation, which determines the short-wavelength
behavior of the structure factor. In Sec.~III we analyze the frequency
dependence of the magnetoconductivity for a short-range disorder
potential. We show that, for nonzero frequencies but $\omega\ll
\omega_c$, the conductivity in quantizing fields becomes a
nonmonotonic function of $B$. In Sec.~IV and the Appendix we develop
an appropriate diagrammatic technique and study the far frequency
tail of the magnetoconductivity due to multiple electron
scattering. In Sec.~V we discuss two types of an intermediate-range
disorder: the smooth disorder potential and the potential of
short-range electron traps; in both cases the conductivity is shown to
be related to diffusion in the electron liquid. Sec.~VI contains
concluding remarks.

\section{Many-electron magnetoconductivity: general arguments}

In the range $\Gamma \gg 1$ the energy of the electron-electron
interaction (EEI)
\begin{equation}
\label{H_ee}
H_{ee}={e^2\over 2}\sum\nolimits_{n\neq m}|{\bf r}_n-{\bf r}_m|^{-1}
\end{equation}
is the largest in the electron system. Therefore even where electrons
do not form a crystal, $\Gamma < \Gamma_W$, electron positions ${\bf
r}_n$ are still correlated. The EEI does not change the total momentum
of the 2DES, and thus does not directly affect the long-wavelength
conductivity $\sigma_{xx}(\omega)$ (the Kohn theorem). However,
momentum transfer from electrons to defects depends on electron
motion, and so $\sigma_{xx}(\omega)$ is ultimately determined by the
EEI.

A standard approach to calculating $\sigma_{xx}$ is based on finding
elementary excitations in the many-electron system and then studying
their scattering by a disorder potential. This approach is not of much
help in the case of a nondegenerate electron liquid, because
elementary excitations are not known \cite{dust}. However, for the
types of disorder that we are interested in, the frequency-dependent
conductivity is determined by electron motion on either short or long
times. This motion can be described even when elementary excitations
are not known, as explained in Appendix~\ref{sec:liquid}.

We will consider magnetoconductivity in a quantizing magnetic field
$B$ applied normal to the electron layer, $\exp(\hbar\omega_c/k_BT)\gg
1$. Then the electron wave function is a wave packet. Its typical size
is the magnetic length $l=(\hbar /m\omega_c)^{1/2}$. The further
analysis is based on the observation \cite{Dykman-All} that, in
addition to a magnetic field, an electron is driven by an electric
field $\ef$ from other electrons. This field is due to electron
density fluctuations, see Appendix~\ref{sec:liquid}. It leads to a
semiclassical drift of the electron wave packet with a group velocity
$cE_{\rm f}/B$.

The semiclassical approximation applies for sufficiently high
temperatures,
\begin{equation}
\label{Trange}
k_BT\gg \hbar\Omega,\;
\Omega = \omega_p^2/\omega_c\equiv 2\pi e^2n/m\omega_c.
\end{equation}
Here, $\Omega$ is the typical frequency of vibrations of the electrons
about their quasi-equilibrium positions in the electron liquid
($\omega_p$ is the plasma frequency for $B=0$, $\Omega\ll
\omega_c$). The picture of moving wave packets, with continuous energy
spectrum, is qualitatively different from the single-electron picture
where the electron energy spectrum is a set of discrete degenerate
Landau levels.

Electron motion leads to averaging of the disorder potential.
Together with inter-electron energy exchange it eliminates
single-electron localization by an arbitrarily weak random potential
studied in the QHE theory
\cite{Pruisken-localization,Huckestein-95}. A typical electron energy
in the liquid is $k_BT$. Therefore a sufficiently strong disorder is
needed in order to localize an appreciable portion of electrons,
potentially leading to a glass transition. In this paper we assume
that the disorder potential is weak and the electron liquid displays
self-diffusion and associated self-averaging.  Specific conditions
depend on the correlation length of the disorder potential and will be
discussed later.

\subsection{Magnetoconductivity for weak disorder potential}

The Hamiltonian of the electron liquid in the presence of disorder has
the form
\begin{eqnarray}
\label{Hamiltonian}
&&H=H_0+ H_{ee} + H_i,\nonumber\\
&&H_i=\sum_{\bf q} V_{\bf q}\,\rho_{\bf q},\quad
\rho_{\bf q}=\sum_n \exp\left(i{\bf
  qr}_n\right).
\end{eqnarray}
Here, $H_0$ is the sum of the single-particle Hamiltonians ${\bf
p}_n^2/2m$ [with ${\bf p}_n=-i\hbar\bm{\nabla}_n+(e/c){\bf A}({\bf
r}_n)$]; $H_{ee}$ is the EEI Hamiltonian and is given by
Eq.~(\ref{H_ee}), and $V_{\bf q}$ are the Fourier components of the
disorder potential.

The long-wavelength magnetoconductivity is given by the correlator of
the total electron momentum ${\bf P}=\sum {\bf p}_n$. The latter
satisfies the equation of motion
%
\[{dP_{\mu}\over dt}=\omega_c\epsilon_{\mu\nu}P_{\nu} -
i \sum\nolimits_{\bf q}q_{\mu}V_{\bf q}\,\rho_{\bf q}\]
($\epsilon_{\mu\nu}$ is the antisymmetric permutation tensor).  The
low-frequency conductivity, $\omega\ll \omega_c$, is determined by
slow time variation of ${\bf P}$. Therefore the time derivative in
this equation can be ignored. The expression for ${\bf P}$ can be then
substituted into the Kubo formula for the conductivity, giving
$\sigma_{xx}(\omega)$ in terms of the correlator of the density
operators $\rho_{\bf q}$ weighted with the disorder potential,

\begin{eqnarray}
\label{eq:cond_iterated}
\sigma_{xx}(\omega)=&&- {e^2l^4[1-\exp(-\beta
  \omega)]\over 4 \hbar^3\omega S} \,
  \int_{-\infty}^{\infty} dt\,   e^{i\omega t}\nonumber\\
&&\times  \sum_{{\bf q},{\bf q}'}\left({\bf q}\,{\bf q}'\right)
 \bigl\langle V_{\bf q}
  V_{{\bf q}'}\,\rho_{\bf q}(t)\rho_{{\bf q}'}(0)\bigr\rangle.
\end{eqnarray}
Here, $\langle\,\cdot\,\rangle$ implies thermal averaging followed by
averaging over realizations of the random potential, $S$ is the area
of the system, and $\beta=\hbar/k_BT$.

In the case of a weak disorder potential, the density-density
correlator in Eq.~(\ref{eq:cond_iterated}) can be evaluated to zeroth
order in $V_{\bf q}$ (the criteria are discussed below). Then the
conductivity can be expressed in terms of the dynamical structure
factor of the electron liquid
\begin{eqnarray}
\label{eq:structure_factor}
{\cal S}({\bf q},\omega)=\int\nolimits_{-\infty}^{\infty}
dt\,e^{i\omega t}\tilde{\cal S}({\bf q},t),\nonumber \\
\tilde{\cal S}({\bf
q},t) = N^{-1}\bigl\langle\rho_{\bf q}(t)\rho_{-\bf q}(0)
\bigr\rangle_0,
\end{eqnarray}
where $N=nS$ is the total number of electrons, and the subscript $0$
means that the correlator is evaluated in the absence of disorder.

For a nondegenerate liquid, it is convenient to write the conductivity
(\ref{eq:cond_iterated}) in the form of an Einstein-type relation
%
\begin{equation}
\label{Einstein}
\sigma_{xx}(\omega)={ne^2D_{s}\over k_BT},\quad
D_{s}=l^2\tau^{-1}(\omega)/4.
\end{equation}
Here, $D_{s}$ can be thought of as a coefficient of electron diffusion
in the disorder potential; it should not be confused with the
coefficient of self-diffusion in the electron liquid discussed in the
Appendix~\ref{sec:liquid}).
The characteristic diffusion length in the disorder
potential is given by the size of the electron wave packet $l$, and
the collision rate is
\begin{equation}
\label{tau}
\tau^{-1}(\omega)={1-e^{-\beta\omega}\over \beta\omega}\,\hbar^{-2}l^2
\sum_{\bf q}q^2\overline{|V_{\bf q}|^2}{\cal S}(q,\omega),
\end{equation}
where the overline denotes averaging over realizations of disorder.

The rate $\tau^{-1}$ (\ref{tau}) is quadratic in the disorder
potential, as in the standard Drude approximation. The expression for
the conductivity (\ref{Einstein}), (\ref{tau}) goes over into the
Drude formula~(\ref{eq:drude}) if one sets $\tau^{-1}=4\tau_{\rm
Dr}^{-1}/ \beta\omega_c $.  However, in contrast to the
single-electron Drude approximation, the dynamic structure factor in
$\tau^{-1}$ is determined by the electron-electron interaction. In
particular ${\cal S}({\bf q},\omega)$ depends on the electron density
$n$.

The factor $[1-\exp(-\beta\omega)]/\beta\omega$ in Eq.~(\ref{tau}) is
equal to 1 in the most interesting frequency range $\beta\omega \ll
1$, which includes the central part of the peak of the low-frequency
conductivity. The frequency dependence of the effective scattering
rate and the conductivity in this range is determined by ${\cal
S}(q,\omega)$. On the other hand, in the analysis of the conductivity
tail we will be interested primarily in the exponent, whereas
$[1-\exp(-\beta\omega)]/\beta\omega$ leads to a smooth frequency
dependence of the prefactor ($\propto \omega^{-1}$ for $\beta\omega\gg
1$). Therefore we omit this factor in what follows.

\subsection{The single-site approximation}
\label{sec:single-site}

The expression (\ref{tau}) is significantly simplified in the
important and most common situation where the correlation length of
the random potential $r_{\rm c}$ is small compared to the
inter-electron distance $n^{-1/2}$. Here, at most one electron is
scattered by a given fluctuation of the potential, for example, by an
impurity in the case of electrons in semiconductors or a
short-wavelength ripplon in the case of electrons on helium. Since the
2DES is strongly correlated, all other electrons are far away.

The condition $r_{\rm c} \ll n^{-1/2}$ allows us to single out the
most important terms in the structure factor ${\cal S}({\bf
q},\omega)$.  The major contribution to the sum over ${\bf q}$ in
Eq.~(\ref{tau}) comes from $q\sim \min(l^{-1}, r_{\rm c}^{-1}) \gg
n^{1/2}$. On the other hand, $\tilde{\cal S}({\bf q},t)$ as given by
Eq.~(\ref{eq:structure_factor}) is a double sum of $\exp[i{\bf
qr}_m(t)]\exp[-i{\bf qr}_{m'}(0)]$ over the electron numbers $m,m'$.
The terms with $m\neq m'$ are rapidly oscillating for $q\gg n^{-1/2}$.
Therefore, when calculating $\tau^{-1}$, one should keep only diagonal
terms with $m=m'$.  We call this the single-site approximation,
\begin{eqnarray}
\label{eq:single_site}
\tilde{\cal S}({\bf q},t)&\approx &\tilde{\cal S}_{\rm ss}({\bf
q},t), \quad q\gg n^{1/2},\nonumber\\
\tilde{\cal S}_{\rm ss}({\bf
q},t)&=& \bigl\langle e^{i{\bf qr}(t)}
e^{-i{\bf qr}(0)}\bigr\rangle_0,
\end{eqnarray}
where ${\bf r}\equiv {\bf r}_m$ stands for the coordinate of an $m$th
electron. The result is independent of $m$, and therefore we
disregarded the electron number in Eq.~(\ref{eq:single_site}).
Respectively, ${\cal S}({\bf q},\omega)\approx {\cal S}_{\rm ss}({\bf
q},\omega)$, where ${\cal S}_{\rm ss}({\bf q},\omega)$ is the Fourier
transform of $\tilde{\cal S}_{\rm ss}({\bf q},t)$.

The transition from the Kubo formula (\ref{eq:cond_iterated}) to
Eqs.~(\ref{Einstein}) - (\ref{eq:single_site}) corresponds to the
approximation of independent scattering events for each individual
electron. It is similar to the standard ladder approximation of the
single-electron theory and applies provided the duration of a
collision $t_{\rm col}$ is small compared to the reciprocal rate of
electron scattering by the disorder potential. In turn, $t_{\rm col}$
is the typical time range that contributes to ${\cal S}_{\rm ss}({\bf
q},\omega)$.

\section{Frequency dispersion of the conductivity: a short-range
  potential}
\label{sec:short-range-potential}

We will first consider the case of a short-range potential with
correlation length
\begin{equation}
\label{short-range}
r_{\rm c}\ll \delta_{\rm f} \equiv(k_BT/m\omega_p^2)^{1/2}.
\end{equation}
Here, $\deltaf$ is the typical thermal displacement of an electron
{}from its quasi-equilibrium position in the electron liquid. As
explained in Appendix~\ref{sec:liquid}, electron motion on distances
smaller than $\deltaf$ and for times much smaller than $\Omega^{-1}$
is a transverse drift in a nearly uniform fluctuational electric field
$\ef$ \cite{AA-85} [the electron vibration frequency $\Omega$ is given
by Eq.~(\ref{Trange})].

We will calculate the structure factor (\ref{eq:single_site}) using a
formulation that differs from the one used in
Refs.~\onlinecite{Dykman-All}. It is advantageous in that it can be
generalized to the case of multiple scattering by disorder potential,
as shown in Sec.~IV.

Thermal averaging of a single-electron operator over the states of the
many-electron system in Eq.~(\ref{eq:single_site}) can be done in two
steps. First we average over the states of a given ($m$th) electron
for a given many-electron configuration. Configuration averaging is
done next. It comes to integration over the relative positions of the
guiding centers ${\bf R}_{m'}$ of all other electrons ($m'\neq m$)
with respect to ${\bf R}_m$. The integration has to be done with the
Boltzmann weighting factor $\exp(-H_{ee}/k_BT)$ [see
Eq.~(\ref{Boltzmann})]. This is because the electron kinetic energies
are eliminated by the Landau quantization, and the only relevant
energy of the system is the potential energy of the electron-electron
interaction.


The first averaging means taking a trace of a corresponding
single-electron operator on the single-electron wave functions of an
($m$th) electron $\psi_k({\bf r})$ (${\bf r}\equiv {\bf r}_m$). The
functions $\psi_k$ belong to the lowest Landau level (LLL) and should
be found assuming that the electron is in a uniform electric field
$\ef$ created by other electrons. No extra weighting factor (in
particular, no Boltzmann factor) has to be incorporated when
calculating the trace. The energy is determined only by the
many-electron configuration, and thermal averaging is done over such
configurations.

The wave functions $\psi_k({\bf r})$ depend on the
many-electron configurations only in terms of the fluctuational field
${\bf E}_{\rm f}$. Therefore the configuration averaging is reduced to
averaging over the distribution of the field $\ef$
(\ref{field_distribution}), which we denote by $\langle\,
\cdot\,\rangle_{\ef}$ (this notation includes subsequent averaging
over realizations of the random potential, if necessary). Overall, the
average value of a single-electron operator $\mathcal O({\bf r})$ can be
written as
\begin{eqnarray}
\label{averaging_short_range}
&&\langle {\mathcal O}({\bf r})
\rangle = {2\pi l^2\over  S}
\Bigl\langle\sum_k \langle\psi_k({\bf r})\vert
\mathcal{O}({\bf r})
\vert \psi_k({\bf r})\rangle\,\Bigr\rangle_{\ef},
\end{eqnarray}
where the prefactor 
is just the reciprocal number of states
of the lowest Landau level.

In order to find the structure factor $\tilde{\cal S}_{\rm ss}({\bf
q},t)$ [Eq.~(\ref{eq:single_site})] to zeroth order in the random
potential we will use the explicit form of the LLL wave functions
$\psi_{k}^{(0)}({\bf r})$ of an electron in the field $\ef$
\begin{equation}
\label{psi}
\psi_{k}^{(0)}({\bf r}) 
={1\over ( L_y l)^{1/2}\pi^{1/4}}
\exp\Biglb(iky-{1\over 2l^2}
\left(x-kl^2\right)^2\Bigrb).
\end{equation}
Here, we chose the $x$-axis in the direction of $\ef$, i.e., $\ef =
E_{\rm f}\,\hat{\bf x}$; $L_y$ is the size of the system in the
$y$-direction, and the coordinate $x$ is counted off from $-eE_{\rm
f}/ m\omega_c^2$. The magnetic field is chosen along the negative
$z$-direction, ${\bf B}=-B\hat{\bf z}$.

The corresponding electron energy
$\varepsilon^{(0)}_k$ (counted off from $\hbar\omega_c/2
-e^2\ef^2/2m\omega_c^2)$) is
\begin{equation}
\label{energy}
\varepsilon^{(0)}_k= eE_{\rm f} kl^2.
\end{equation}

The structure factor
(\ref{eq:single_site}) is determined by the trace of a product of the
single-electron operators $\exp\biglb(\pm i{\bf qr}(t)\bigrb)$ taken
at different
times in the Heisenberg representation. From Eqs.~(\ref{psi}),
(\ref{energy}) we have
\begin{equation}
\label{mat_el}
\bigl\langle\psi_k^{(0)}\bigl\vert e^{i{\bf qr}(t)} e^{-i{\bf qr}(0)
  }
\bigr\vert\psi_k^{(0)}\bigr\rangle 
 \approx
e^{-q^2l^2/2}\,
e^{it {\bf qv}_D },
\end{equation}
where ${\bf v}_D\equiv c\ef\times {\bf B}/B^2$ is the semiclassical
drift velocity; for chosen axes the vector ${\bf v}_D$ points in the
$y$-direction. In Eq.~(\ref{mat_el}) we disregarded fast-oscillating
terms $\propto \exp(\pm in\omega_ct)$ with $n\geq 1$. Such terms make
extremely small contribution to the conductivity for $\omega\ll
\omega_c$.

Following the procedure (\ref{averaging_short_range}), in order to
find $\tilde{\cal S}_{\rm ss}({\bf q},t)$ we have to average the
right-hand side of Eq.~(\ref{mat_el}) over the fluctuational field
$\ef$. For the Gaussian\cite{Fang-Yen-97} field distribution
(\ref{field_distribution})
 this gives
\begin{equation}
\label{correlator_0}
\tilde {\cal S}_{\rm ss}({\bf q},t)= \exp\Biglb(-{1\over
2}q^2l^2\Bigl(1+ \frac{ t^2}{\tilde t_e^2}\Bigr)\Bigrb).
\end{equation}
Eq.~(\ref{correlator_0}) shows that the structure factor decays very
fast for wave numbers $q\gg 1/l$. For $ql\sim 1$ it also rapidly
decays with time for $t\gg \tilde t_e$.

The characteristic time $\tilde t_e$ is simply related to the
r.m.s. drift velocity and the r.m.s. fluctuational field $\langle
\ef^2\rangle$,
\begin{equation}
\label{tildet_e}
\tilde t_e= {\sqrt 2 l\over \langle {\bf v}_D^2\rangle^{1/2}}
= { \sqrt 2\hbar\over el \langle\ef^2\rangle^{1/2}}
\end{equation}
(the choice of the coefficients is convenient for
Eq.~(\ref{sigma_explicit}) below).  A closely related time
\begin{equation}
\label{t_e}
t_e=l(B/c)\,\langle E_{\rm f}^{-1}\rangle  \sim (\Omega\, k_BT/\hbar)^{-1/2}.
\end{equation}
was introduced previously \cite{Dykman-All} as the average time of
flight of an electron wave packet over a distance $l$ in the crossed
fields $\ef$, ${\bf B}$.  In the case of a short-range disorder
potential ($r_{\rm c} \ll l$), $t_e$ is the duration of an electron
collision with the fluctuation of the potential (a point defect).  For
a Gaussian distribution of the fluctuational field assumed here we
have $\tilde t_e=(2/\pi)^{1/2} t_e$.

Eqs.~(\ref{tau}), (\ref{correlator_0}) give the conductivity in a
simple form.  The typical values of ${\bf q}$ transferred in an
electron collision are $q_{\rm c}= (r_{\rm c}^2+l^2)^{-1/2}$. The
duration of a collision is the time over which the correlator
(\ref{correlator_0}) decays, $t_e/q_{\rm c}l$. For a short-range potential
(\ref{short-range}), we have $t_e/q_{\rm c}l \ll \Omega^{-1}$. This
justifies the assumption that the fluctuational field remains constant
during a collision, which is equivalent to the assumption that $t\ll
\Omega^{-1}$.

The condition for the disorder potential to be
weak so that collisions occur successively in time is
\begin{equation}
\label{weak-scattering}
t_e\left[(r_{\rm c}/l)^2+1\right]^{1/2}\ll \tau(0),
\end{equation}
where $\tau^{-1}(\omega)$ is given by Eq.~(\ref{tau}) and is quadratic
in the potential strength.

The typical frequency $\Omega$ [Eq.~(\ref{Trange})] of electron vibrations
about its quasi-equilibrium position is also the rate of
inter-electron energy exchange. The time interval between successive
collisions of a vibrating electron with the same short-range scatterer
$\sim (q_{\rm c}\deltaf)^2\Omega^{-1}$ is much larger than
$\Omega^{-1}$. Therefore an electron looses coherence between
successive collisions. This shows that interference effects leading to
weak localization in the single-electron approximation are not
important, for weak scattering.

\subsection{The conductivity for a $\delta$-correlated random potential}

The expression for the conductivity can be obtained in an explicit
form in the important case of a $\delta$-correlated random potential
$\overline{V({\bf r})V({\bf r'})}=v^2\delta({\bf r} - {\bf r}')$, or
\begin{eqnarray}
\label{delta_corr}
\overline{|V_{\bf q}|^2} =\pi \hbar^2\gamma^2l^2/2S,\quad
\hbar\gamma=(2/\pi)^{1/2}v/l.
\end{eqnarray}
The parameter $\gamma\propto 1/l$ introduced here is a convenient
characteristic of the random potential in the problem of electrons in
a quantizing magnetic field. It gives the width of the peak of the
density of states $\rho(E)$ in the single-electron approximation
\cite{Wegner-83}; in particular, on the tail $\rho(E)\propto
\exp(-4E^2/\hbar^2\gamma^2)$, see Ref.~\onlinecite{Ioffe-81}. From
Eq.~(\ref{tau}), the scattering rate for the potential
(\ref{delta_corr}) is $\tau^{-1}(0)\sim \gamma^2 t_e$, and the
condition for the random potential to be weak takes a simple form
\[\gamma \ll t_e^{-1}.\]

Collecting Eqs.~(\ref{Einstein}), (\ref{tau}), (\ref{correlator_0}),
(\ref{delta_corr}) we obtain a simple explicit expression for the
frequency-dependent magnetoconductivity,
\begin{eqnarray}
\label{sigma_explicit}
\sigma_{xx}(\omega)&\approx&{\pi\over16}\,{n e^2\beta\gamma^2\tilde t_e\over
m\omega_c}\,\sigma_1(\omega)\nonumber \\
\sigma_1(\omega)&=&\left( 1+{\omega
\tilde t_e}\right)e^{-\omega\tilde t_e}.
\end{eqnarray}
In the quasi-static limit of small $\omega t_e$
Eq.~(\ref{sigma_explicit}) coincides with our previous result
\cite{Dykman-All}. The conductivity $\sigma_{xx}(0)$ has a form of a
single-electron conductivity in a magnetic field, with a scattering
rate $\tau^{-1}\sim \gamma^2t_e$ quadratic in the disorder
potential. However, in contrast to $\tau_{\rm Dr}^{-1}$
Eq.~(\ref{eq:drude}), the value of $\tau^{-1}$ is fully determined by the
EEI.  Through the factor $t_e$ it depends on the fluctuational
electric field that drives an electron during a collision with a
scatterer. It scales with the density of the electron liquid as
$n^{-3/4}$, so that the overall conductivity $\sigma_{xx}(0)\propto
n^{1/4}$.

\begin{figure}[ht]
    \includegraphics[width=3.00in]{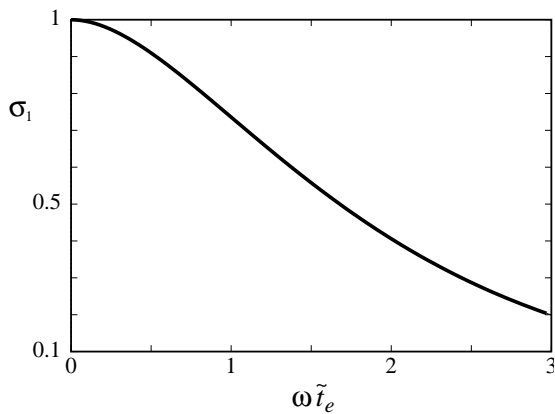}
\caption{ The frequency dependence of the many-electron conductivity
  for short-range disorder as given by Eq.~(\protect\ref{sigma_explicit}).
}
\label{fig:weak_dis}
\end{figure}

The frequency dependence of the conductivity is determined by the
dimensionless function $\sigma_1(\omega)$, which is shown in
Fig.~\ref{fig:weak_dis}.  It peaks at zero frequency and
monotonically decays with increasing $\omega$. In contrast to the
Drude conductivity in the absence of a magnetic field, which has a
Lorentzian peak $\sigma_{xx}\propto 1/(1+\omega^2\tau^2)$, the peak of
$\sigma_1$ is strongly non-Lorentzian. The characteristic width
$t_e^{-1}$ of the peak of $\sigma_1$ is {\it independent} of the
disorder potential. Its dependence on the electron density,
temperature, and the magnetic field is of the form $t_e^{-1}\propto
n^{3/4}T^{1/2}B^{-1/2}$.

The tail of the peak is exponential in $\omega$. Interestingly, it has
the exact form of the Urbach rule \cite{Urbach-rule}, i.e.,
$|\ln\sigma_1|\propto \omega$.  [However, for very large $\omega t_e$
where multiple scattering becomes important, this form is modified,
see Sec.~\ref{sec:conductivity-tail}.]  The shape of the tail can be
understood by noticing that the conductivity is formed by processes in
which the {\em energy\/} $\hbar\omega$ of absorbed photons goes to the
many-electron system. In an individual absorption process the involved
electron moves by the distance $\delta R=\hbar\omega/e\,|\ef|$ along
the fluctuational electric field $\ef \parallel \hat x$.  The squared
matrix element of a dipolar electron transition accompanied by a
displacement $\delta {\bf R}\equiv \delta R\,\hat{\bf x}$ is
determined by the (squared) overlap integral of the wave functions
(\ref{psi})
\begin{equation}
 \label{eq:overlap}
  \vert\langle\psi_k^{(0)}({\bf r})\vert\psi_k^{(0)}({\bf r+\delta
    R})\rangle\vert^2= \exp\left[-(\delta
    R)^2/2l^2\right].
\end{equation}

On the other hand, the probability (\ref{field_distribution}) to
have a fluctuational field $E_{\rm f} = \hbar\omega/(e\,\delta R)$ is
\[p\propto \exp\bigglb(-{(\hbar\omega)^2\over e^2(\delta
R)^2\,\langle \ef^2\rangle}\biggrb).\]

By optimizing the product of the two exponentials with respect to
$\delta R$, we obtain that the conductivity is $\propto \exp(-\omega
\tilde t_e)$, with account taken of the expression (\ref{tildet_e})
for $\tilde t_e$. This is in agreement with Eq.~(\ref{sigma_explicit}).

It is important to check the assumption that the field $\ef$ is
uniform over relevant distances. In a strongly correlated system
$|{\bm\nabla}\ef| \sim e\,n^{3/2}$. Therefore for optimal $\delta R$ and
$E_{\rm f}$ the relative change of the field on the
distance $\delta R$ is
$$
{|{\bm \nabla}\ef|\,\delta R\over E_{\rm f}} \sim {e\,n^{3/2}\delta
R\over E_{\rm f}} ={e^2n^{3/2}(\delta R)^2\over \hbar\omega} \sim
\Omega t_e\ll 1.$$ It is interesting that this condition does not
impose limitations on $\omega$.

Another interesting feature of the many-electron microwave
conductivity $\sigma_{xx}(\omega)$ is its nonmonotonic dependence on
the magnetic field. Since $\gamma\propto 1/l \propto B^{1/2}$ [see
Eq.~(\ref{delta_corr})] and $t_e\propto B^{1/2}$ [Eq.~(\ref{t_e})],
the static conductivity $\sigma_{xx}(0)\propto B^{1/2}$ is {\it
increasing} with $B$ for quantizing fields. This happens because, as
$B$ increases, the electron wave function becomes more localized, thus
increasing the effective strength of coupling to short-range
scatterers. At the same time, the electron drift velocity in the
fluctuational field decreases, and as a result the characteristic
collision duration $t_e$ increases with $B$, leading to the overall
scattering rate $\tau^{-1}\sim \gamma^2t_e \propto B^{3/2}$
\cite{Dykman-All}. The increase of $\sigma_{xx}(0)$ with increasing
$B$ has been confirmed experimentally \cite{Lea-97}.

\begin{figure}[htbp]
  \begin{center}
    \includegraphics[width=3.0in]{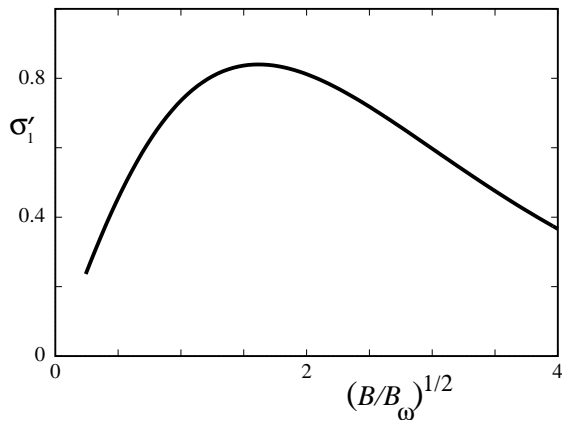}
\end{center}                               
\caption{The dependence of the reduced microwave many-electron
conductivity $\sigma'_1(\omega) = \omega \tilde t_e \sigma_1(\omega)$
on the magnetic field for weak $\delta$-correlated random potential
and for $k_BT\gg \hbar\omega$.  Both the characteristic scaling
magnetic field $B_{\omega}=B/(\omega \tilde t_e)^2$ and the ratio
$\sigma_{xx}(\omega)/\sigma'_1(\omega)$ are independent of $B$, see
Eqs.~(\protect\ref{tildet_e}) and (\ref{sigma_explicit}).}
\label{fig:m_e_Bpeak}
\end{figure}

For nonzero frequencies, $\sigma_{xx}(\omega)$ displays a peak as a
function of $B$, see Fig.~\ref{fig:m_e_Bpeak}. For comparatively small
(but still quantizing) fields we have $\omega t_e\ll 1$, and then
$\sigma_{xx}(\omega)\propto t_e\propto B^{1/2}$, as in the static
limit. On the other hand, for $B$ such that $\omega t_e\gg 1$, the
conductivity falls down exponentially with increasing $\omega
t_e$. The position $B_{\max}$ of the peak of the conductivity is given
by the Golden ratio
\begin{equation}
  \label{m_e_max}
  (\omega \tilde t_e)_{\max} = {1+\sqrt{5}\over2},\quad
  B_{\max}\propto\omega^{-2}.
\end{equation}

We note that the microwave magneto-conductivity displays a peak as a
function of $B$ in the single-electron approximation as well
\cite{Kuehnel01}. However, the shape of the peak is totally
different. In particular, the high-$B$ decay of the single-electron
$\sigma_{xx}(\omega)$ is related to localization of electron states in
the random potential and is described by a power law. Many-electron
effects lead to delocalization, and the decay of the conductivity with
$B$ becomes exponential, cf. Fig.~\ref{fig:m_e_Bpeak}. Of course, for
very strong $B$ the approximation of weak random potential $\gamma
t_e\ll 1$ breaks down, and then the decay of $\sigma_{xx}(\omega)$ with $B$
slows down.

We note also that in the single-electron theory $\sigma_{xx}(0)=0$ due
to electron
localization\cite{Wang-Fisher-Girvin-Chalker,Kuehnel01}. Therefore
$\sigma_{xx}(\omega)$ as function
of frequency has a peak at nonzero frequency which depends on
$B$. This is in contrast with the monotonic decay of
$\sigma_{xx}(\omega)$ in the many-electron theory shown in
Fig.~\ref{fig:weak_dis}.

\section{Conductivity tail: the effect of multiple scattering}
\label{sec:conductivity-tail}

It follows from the results of Sec.~III [see Eq.~(\ref{psi})] that
electron wave functions in crossed electric and magnetic fields
display Gaussian decay along the electric field. One may expect that
multiple scattering of an electron in a short-range random potential
will slow down this decay, as it does for localized electrons in the
absence of an electric field \cite{Thouless}. In turn, a slower
spatial decay may lead to a slower decay of the many-electron
conductivity $\sigma_{xx}(\omega)$ with frequency on the tail where
$\omega t_e\gg 1$. This is because it becomes more probable for an
electron to shift by a larger distance along the fluctuational
electric field $\ef$ and therefore to absorb a photon with higher
energy.

The approximation of an electron moving in a static uniform field
$\ef$ is well justified in the frequency range $\omega t_e\gg
1$. Indeed, the many-electron field $\ef$ changes over time $\sim
\Omega^{-1}$, see Appendix~\ref{sec:liquid}. It remains constant over
the duration $\omega^{-1}$ of absorption of a photon, because
$\omega^{-1}\ll t_e \ll \Omega^{-1}$. The argument in favor of the
field uniformity is based on the fact that the absorption tail is
formed by large fields $\ef$. They are experienced by electrons that
are far away from their quasi-equilibrium positions. The larger is the
field the larger the distance to the quasi-equilibrium position should
be. On the other hand, this distance is also the scale on which the
field is spatially nonuniform. We will see that it will largely exceed
the electron displacement $\hbar\omega/eE_{\rm f}$ during
absorption. Both lengths will be assumed much smaller than the
inter-electron distance $n^{-1/2}$.

A large electron displacement requires a large momentum transfer to
the random potential, $q\gg n^{-1/2}$. Therefore it is a good
approximation to evaluate the correlator in the expression for the
conductivity (\ref{eq:cond_iterated}) using the single-site
approximation. This means that the product $\rho_{\bf
q}(t)\rho_{{\bf q}'}(0)$ in Eq.~(\ref{eq:cond_iterated}) should be replaced by
$\sum\nolimits_m\exp[i{\bf qr}_m(t)]\exp[i{\bf q}'{\bf r}_m(0)]$. Then
\begin{eqnarray}
\nonumber
  \sigma_{xx}(\omega)&=&-{ne^2  l^4\over 4k_B T\,\hbar^2}
  \int_{-\infty}^\infty dt\,e^{i\omega t}\\
&& \times\sum_{{\bf q,q}'}({\bf qq}')
  \left\langle V_{\bf q} V_{{\bf q}'}\,
  e^{i{\bf qr}(t)}   e^{i{\bf q'r}(0)}\right\rangle,
  \label{cond-iterated1}
\end{eqnarray}
where ${\bf r}\equiv {\bf r}_m$ [the result is independent of the
electron number $m$]. For $\hbar\omega \agt k_BT$ we should replace
$(k_BT)^{-1}$ with $[1-\exp(-\beta\omega)]/\hbar\omega$; however, as
noted above, it will only affect the prefactor in the conductivity.

The averaging in Eq.~(\ref{cond-iterated1}) can be done following the
prescription (\ref{averaging_short_range}), i.e., one first calculates
a trace over the single-electron wave functions $\psi_k$ in a
fluctuational field and then averages over the field. In contrast to
the calculation in Sec.~III, to allow for multiple scattering one
should use wave functions found with account taken of the disorder
potential $V({\bf r})$. In our approximation averaging over
realizations of $V({\bf r})$ and $\ef$ is done independently. It turns
out to be more convenient to average over $V({\bf r})$ first.

In what follows the random potential is assumed to be Gaussian and
$\delta$-correlated, with the correlator (\ref{delta_corr}). We will
use the Green function technique. In contrast to what was done in the
analysis of the tail of the wave function \cite{Thouless}, this
technique has to be formulated in the frequency domain. We will show
that this leads to a somewhat unusual set of diagrams.

\subsection{The projected Green function}

In order to find the low-frequency conductivity it is convenient to
use the Green function $G_{\varepsilon}({\bf r,r}')$ ``projected'' on
the lowest Landau level. It is constructed from the LLL wave functions
$\psi_k({\bf r})$ of an electron in the random potential $V({\bf r})$
and in the electric field $\ef$,
\begin{equation}
\label{Green-general}
G_\varepsilon({\bf r,r}') = \sum\nolimits_k \psi_k({\bf
r})\psi_k^*({\bf r}')(\varepsilon - i0-
\varepsilon_k)^{-1},
\end{equation}
where $\varepsilon_k$ is the single-electron energy of the state
$k$. The wave functions $\psi_k$ are linear combinations of the LLL wave
functions $\psi_k^{(0)}$ [Eq.~(\ref{psi})] in the absence of disorder.
Following the averaging procedure (\ref{averaging_short_range}), we
obtain from Eq.~(\ref{cond-iterated1})
\begin{eqnarray}
  \label{sigma_Green}
\lefteqn{\!\!\!\!\!\!\sigma_{xx}(\omega)= {ne^2
  l^4\over 4 k_BT\hbar }\,{l^2\over S}\,{\rm
  Re}\sum_{\nu=x,y}\int_{-\infty}^{\infty} d\varepsilon\int d{\bf
  r}\,d{\bf r}'}\nonumber\\
&&\!\!\!\!\times\bigl\langle V({\bf r})V({\bf
  r}')\,\partial_{r_{\nu}}\partial_{r_{\nu}}\, G_\varepsilon({\bf
  r,r}') G^*_{\varepsilon+\hbar\omega}({\bf
  r,r}')\bigr\rangle_{\ef}.
\end{eqnarray}
%

We will consider the random potential $V({\bf r})$ as a
perturbation. The zeroth-order Green function $G^{(0)}_\varepsilon$
can be found using the explicit expressions (\ref{psi}),
(\ref{energy}) for the wave functions and the energy in the absence of
disorder,
\begin{eqnarray}
  \label{g_representation}
  &&\!\!\!\!\!\!
G_\varepsilon^{(0)}({\bf r},{\bf r}')=\pi^{-1/2}l\,g({\bf r},{\bf
  r}') \,
\int dk\,\nonumber \\
&&\!\!\times
  {\exp\biglb(-[2l^2k-(x+x')-i(y-y')]^2/4l^2\bigrb)\over \varepsilon -e
  E_f l^2 k-i0},
\end{eqnarray}
where the $x$-axis is chosen along the field $\ef$ as in
Eq.~(\ref{psi}).

The function
\begin{equation}
  \label{g_function}
  g({\bf r,r}')={1\over 2\pi l^2}
e^{-({\bf r}-{\bf r}')^2/4l^2}\,e^{i(x+x')(y-y')/2l^2}
\end{equation}
in Eq.~(\ref{g_representation}) has a simple meaning. It gives the
(minus) right-hand side of the Schr\"odinger equation for the
projected Green function $G_{\varepsilon}({\bf r,r}')$, that is it
replaces the $\delta$-function $\delta({\bf r}-{\bf r}')$ in the
Schr\"odinger equation for a standard Green function. The function
$g({\bf r,r}')$ is localized in a narrow region $|{\bf r}- {\bf
r}'|\lesssim 2l$ and leads to a Gaussian fall-off of
$G_{\varepsilon}^{(0)}$ for large distances.

The full Green function $G_\varepsilon$ is determined by the
Dyson equation.  Its solution can be written symbolically as
\begin{equation}
\label{series}
G_\varepsilon = G_\varepsilon^{(0)} + G_\varepsilon^{(0)}\cdot V\cdot G_\varepsilon^{(0)} + G_\varepsilon^{(0)}\cdot V\cdot
G_\varepsilon^{(0)}\cdot V\cdot G_\varepsilon^{(0)} + \ldots
\end{equation}
where the central dot implies integration over internal coordinates, like
$\int d{\bf r}_iG_\varepsilon^{(0)}({\bf r}_{i-1},{\bf
r}_i)V({\bf r}_i)G_\varepsilon^{(0)}({\bf r}_{i},{\bf
r}_{i+1})$.
We emphasize that, even though the Green function $G_\varepsilon$ is
projected on the LLL, Eq.~(\ref{series}) contains the {\em full}
rather than the projected disorder potential $V({\bf r})$.

A straightforward calculation shows that, to the lowest order in $V$,
the conductivity obtained from Eq.~
(\ref{sigma_Green}), coincides with the result of Sec.~III; in this
approximation the full Green function $G_\varepsilon$ in
Eq.~(\ref{sigma_Green}) has to be replaced with
$G^{(0)}_\varepsilon$.

\subsection{Diagrams for high-frequency conductivity}
\label{sec:diagrams}

According to Eqs.~(\ref{g_representation}), (\ref{g_function}), the
Green function $G^{(0)}_\varepsilon({\bf r,r}')$ is mostly localized
in a narrow region $|{\bf r}- {\bf r}'|\lesssim 2l$. As a function of
energy, it peaks at the (scaled) midpoint in the $\ef$-direction, where
$\varepsilon = eE_{\rm f}(x+x')/2$ for $y= y'$. Near the maximum,
\begin{eqnarray}
\label{Green_approximate}
&&G^{(0)}_\varepsilon({\bf r},{\bf r}') \approx {g({\bf
r,r}')\over \varepsilon-i0 -eE_{\rm f}r^{(c)}}.\\
&&r^{(c)}\equiv r^{(c)}({\bf r,r}') = [(x+x')+i(y-y')]/2\nonumber
\end{eqnarray}

With this in mind, we now consider the expression (\ref{sigma_Green})
for the conductivity in terms of the product of the Green functions
and think of $G_\varepsilon$ and $G_\varepsilon^*$ as given by the
perturbation series (\ref{series}). In the product
of the series we need to find terms
containing $G^{(0)}_\varepsilon({\bf r},{\bf r}')$ and
$[G^{(0)}_{\varepsilon+\hbar\omega}(\tilde{\bf r}, \tilde{\bf r}')]^*$
with
\[eE_{\rm f}[r^{(c)}(\tilde{\bf
r},\tilde{\bf r}')- r^{(c)}({\bf r,r}')]\approx \hbar\omega.\]
Such terms describe absorption of a photon accompanied by an electron
displacement by $\delta R= \hbar\omega/eE_{\rm f}$. The displacement
results from scattering by the random potential.

\begin{figure}[ht]
\includegraphics[width=\columnwidth]{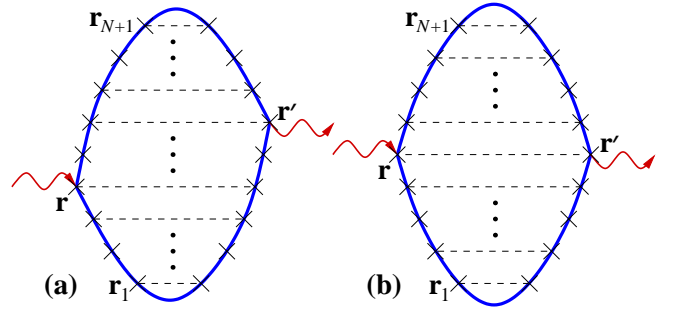}
\caption{Diagrams for the conductivity (\protect\ref{sigma_Green}) in
the coordinate representation. Segments of the solid lines connecting
the points ${\bf r}\equiv{\bf r}_s$ and ${\bf r}'\equiv {\bf r}'_{s'}$
in the upper and lower halfplanes are the Green functions
$G^{(0)}_\varepsilon({\bf r}_i,{\bf r}_{i+1})$ and $
[G^{(0)}_{\varepsilon +\hbar\omega}({\bf r}'_i,{\bf r}'_{i+1})]^*$,
respectively. Crosses (``{\sf x}'') correspond to the Gaussian
$\delta$-correlated potential $V$ at points ${\bf r}_i$ [subsequent
integration over ${\bf r}_i$ is implied]. Dashed lines indicate
averaging over $V$. Wavy lines mark the ``external'' vertices $V({\bf
r})$ and $V({\bf r}')$ in Eq.~(\protect\ref{sigma_Green}) and imply
differentiation of the product of the Green functions over ${\bf
r,r}'$ prior to averaging over $V$. }
\label{fig:fish}
\end{figure}

Graphically, the leading-order contribution to conductivity can be
represented by a sum of the
``fat fish'' diagrams illustrated in Fig.~\ref{fig:fish} in the
coordinate representation. In this figure, the wavy lines mark the
points ${\bf r}$ and ${\bf r}'$ where the derivatives of the product
of the Green functions are taken in Eq.~(\ref{sigma_Green}). The lines
above and below these points represent the series (\ref{series}) for
$G_\varepsilon({\bf r,r}')$ and $[G_{\varepsilon+ \hbar\omega}({\bf
r,r}')]^*$, respectively. The crosses (``{\sf x}'') indicate the
factors $V$ in the expansion (\ref{series}) at points ${\bf r}_i$
where the electron is ``scattered''. Solid lines between the crosses
denote the Green functions $G_\varepsilon^{(0)}({\bf r}_i,{\bf
r}_{i+1})$ and $[G_{\varepsilon+\hbar\omega}^{(0)}({\bf r}'_i,{\bf
r}'_{i+1})]^*$ that describe electron propagation between collisions.

An $N$\,th order diagram has the total number of crosses equal to
$2(N+1)$ (two crosses come from the external vertices in
Fig.~\ref{fig:fish}). We start with the Green functions that
correspond to the top and bottom segments in Fig.~\ref{fig:fish}. Each
of them connects points with equal coordinates, ${\bf r}^{}_{N+1}$ and
${\bf r}_{1}$, respectively. Their product is
\begin{eqnarray}
\label{edge}
&&G^{(0)}_\varepsilon({\bf r}_1,{\bf
r}_1)\,[G^{(0)}_{\varepsilon+\hbar\omega}({\bf r}^{}_{N+1},{\bf
r}^{}_{N+1})]^* \nonumber\\ &&\!\!\!\!\propto
(\varepsilon -i0-eE_{\rm f}x_1)^{-1}(\varepsilon +\hbar\omega
+i0-eE_{\rm f}x^{}_{N+1})^{-1}.
\end{eqnarray}

Upon integration over $\varepsilon$ in Eq.~(\ref{sigma_Green}), the
expression (\ref{edge}) goes over into $2\pi^2\delta\biglb(\hbar\omega -
eE_{\rm f}(x^{}_{N+1}-x_1)\bigrb)$. This is the equation of energy
conservation in a photon-induced transition ${\bf r}_1\to {\bf
r}_{N+1}$. Such a transition is exactly the process that diagrams
in Fig.~\ref{fig:fish} describe.

This picture provides a physical insight into the diagrams. It shows
that the absorption does {\it not} occur in a transition ${\bf r}\to
{\bf r}'$ between the external points of the diagram. Important for
the transition is the admixture of the wave functions centered away
{}from ${\bf r}, {\bf r}'$ (in fact, maximally far away, see below).

The solid lines in Fig.~\ref{fig:fish} other than the top and bottom
segments connect spatially-separated points ${\bf
  r}_i$, ${\bf r}_{i+1}$.  The leading exponential terms in the
dependence of the Green functions on the distance between the points
is given by overlap functions $g({\bf r}_i,{\bf r}_{i+1})$ [see
Eq.~(\ref{Green_approximate})]. The product of the $g$-functions
entering the $N$\,th order diagrams can be written as
\begin{eqnarray}
\label{g_factor}
\tilde g^{}_N(\{{\bf r}_j\})&=& g({\bf r}_1,{\bf
r}_1)\,g({\bf r}^{}_{N+1},{\bf
r}^{}_{N+1})\prod_{i=1}^{N}\left|g({\bf r}^{}_i,{\bf
r}^{}_{i+1})\right|^2\nonumber\\
&\propto& \exp\Biglb(-\sum_i({\bf r}^{}_i-{\bf
r}^{}_{i+1})^2/2l^2\Bigrb).
\end{eqnarray}
with $\{{\bf r}_j\}\equiv {\bf r}_1,\ldots,{\bf r}_{N+1}$. We have
incorporated into Eq.~(\ref{g_factor}) the terms from the ``external''
coordinates ${\bf r}\equiv{\bf r}_s$, ${\bf r}'\equiv {\bf r}'_{s'}$.

For large $\hbar\omega/NeE_{\rm f}l$ the integral over ${\bf
r}_2,\ldots,{\bf r}^{}_{N+1}$ can be evaluated by steepest descent,
with the constraint $\hbar\omega = eE_{\rm f}(x^{}_{N+1}-x_1)$ [the
integral over ${\bf r}_1$ cancels the factor $S^{-1}$ in
Eq.~(\ref{sigma_Green})]. As in the case of underbarrier tunneling in
a magnetic field \cite{Thouless}, the extreme points are equidistant,
\begin{equation}
\label{extreme_points}
x_i^{(e)} = x_1 +
(i-1)\hbar\omega/NeE_{\rm f},\;\; y_i^{(e)}=y^{(e)}_{1}
\end{equation}
for $i=2,\ldots, N+1$.  Then, at the extremum,
%
\begin{equation}
\label{log_g}
\ln\tilde
g^{}_N(\{{\bf r}_j^{(e)}\})\approx
-(\hbar\omega/eE_{\rm f}l)^2/2N.
\end{equation}
This expression has a simple physical meaning. The factor $N$ in the
denominator shows the scattering-induced increase of the overlap
integral between the electron wave functions centered at points ${\bf
r}_1$ and ${\bf r}^{}_{N+1}$, which are separated by
$\hbar\omega/eE_{\rm f}$.

Except for the top and bottom Green functions (\ref{edge}), all other
Green's functions in the diagrams in Fig.~\ref{fig:fish} are
nonresonant.  Their saddle-point values can be evaluated using the
approximate expressions~(\ref{Green_approximate}). The diagrams in
Fig.~\ref{fig:fish}(a) have ${\bf r}_s\equiv {\bf r}$ that differs
from ${\bf r}_{s'}\equiv {\bf r}'$. These diagrams describe
interference of different tunneling paths and can be negative or
positive.  The total of all diagrams in Fig.~\ref{fig:fish} is, of
course, positive. The interference of paths affects only the
prefactor in the conductivity. The leading exponential dependence on
the distance $\hbar\omega/eE_f$ and on the diagram order $N$ is not
affected. It is the same for all diagrams in Fig.~\ref{fig:fish}.
Therefore here we will give results only for the ``diagonal''
terms, which are described by the diagrams in Fig.~\ref{fig:fish}(b).

The $N$\,th -order diagrams in Fig.~\ref{fig:fish}(b) contain a
multiplier $[\pi\hbar^2\gamma^2l^2/2]^{N+1}$ from the intensity of the
random potential (\ref{delta_corr}).  Together with the factors from
the energy denominators [see Eq.~(\ref{Green_approximate})] and the
factors from the integration over $d{\bf r}_i$ around the extreme
points (\ref{extreme_points}), these coefficients give an
$N$-dependent factor
\begin{eqnarray}
\label{combinatorial0}
C_s = [N\gamma/\omega]^{2N}\left[(2s-3)!!\,(2N-2s+1)!!\right]^{-2},
\end{eqnarray}
which also depends on the position $s$ (${\bf r}={\bf r}'\equiv {\bf
  r}_s$) of the wavy lines in
Fig.~\ref{fig:fish}(b) (we assume that $1<s<N+1$). For large $N\gg 1$,
the factor $C_s$ is maximal for $s=N/2$,
\begin{equation}
\label{combinatorial}
\ln C_{\max}\approx 2N\ln(\gamma/\omega)+2N
\end{equation}
The condition $s=N/2$ indicates that photon-induced transitions
preferably occur between the states that are maximally (and equally)
separated from the ``external'' points ${\bf r}, {\bf r}'$. This is
optimal in terms of maximizing the overlap integral of states with
given energy separation.

A more detailed calculation \cite{LP-diagrams} shows that, in order to
allow for compensation from diagrams in Fig.~\ref{fig:fish}(a), it is
necessary to incorporate corrections to the leading-order steepest
descent integrals. However, as we already mentioned, these corrections
do not affect the leading term in $\ln\sigma_{xx}$. We also note that,
for a given $\ef$, the in-plane conductivity becomes anisotropic. This
anisotropy leads to different prefactors for the conductivity in the
directions parallel and perpendicular to $\ef$ \cite{LP-diagrams}.

\subsection{Frequency dependence of the logarithm of the conductivity}

The logarithm of the conductivity $\sigma_{xx}(\omega)$ is given by
the maximal value of the sum of the expressions (\ref{log_g}),
(\ref{combinatorial}) with respect to the order of the diagram $N$. To
average over the fluctuational electric field, one has to add the
logarithm of the field distribution and find the maximum of the
resulting expression over $\ef$. With the Gaussian field distribution
(\ref{field_distribution}), we obtain
\begin{eqnarray}
\label{sigma_asymptotic}
\ln\sigma_{xx}(\omega)\approx -(3/ 2^{1/3})\,(\omega \tilde t_e)^{2/3}
\left[\ln(\omega/\gamma)-1\right]^{1/3},
\end{eqnarray}
where $\tilde t_e$ is given by Eq.~(\ref{tildet_e}).

Eq.~(\ref{sigma_asymptotic}) is the central result of this section. It
shows that the logarithm of the conductivity depends on frequency as
$\omega^{2/3}$ for large $\omega$. This form of decay is a result of
multiple scattering in the random potential, which helps an electron
move along the fluctuational field as it absorbs a photon. The
crossover to Eq.~(\ref{sigma_asymptotic}) from the single-scattering
approximation (\ref{sigma_explicit}) [where $|\ln\sigma_{xx}|\propto
\omega$] occurs when the optimal number of scattering events $N\sim
[\omega t_e/\ln(\omega/\gamma)]^{2/3}$ becomes large. The
corresponding value of $\omega$ depends on both the many-electron
fluctuational field and the intensity of the random potential.

Several comments need to be made about this result.  First, the
diagrams in Fig.~\ref{fig:fish} are unusual for a transport
problem. They are neither the standard ``ladder diagrams'', nor the
maximally crossing diagrams. Rather the diagrams in
Fig.~\ref{fig:fish}(b) are the maximally wrapped (embedded) diagrams
for the self-energy.  Here they appear, because absorption of a photon
is accompanied by only one ``real'' scattering by the random
potential.  The role of multiple scattering is to alleviate the
Gaussian decay of the electron wave function along the fluctuational
field.  Second, as the frequency increases, the probability of a
realization of an optimal fluctuational field $E_{\rm f}$ may become
non-Gaussian. In particular, in the region where $\ln p(E_{\rm f})$ is
nearly linear in $E_{\rm f}$ \cite{Fang-Yen-97}, we obtain
$|\ln\sigma_{xx}| \propto \omega^{1/2}$, i.e., even a slower decay
than Eq.~(\ref{sigma_asymptotic}).

\section{Magnetoconductivity in a smooth random potential}

In many physically interesting cases the correlation length of the
random potential $r_{\rm c}$ is large (or effectively large, see
below) compared to the typical size of the electron wave packet. A
well-known example is provided by electron systems in semiconductor
heterostructures, where much of the disorder potential comes from
the donors that are spatially separated from the 2DES. A sufficiently
weak smooth random potential does not lead to a glass transition in an
electron liquid. The liquid should then display a nonzero static
conductivity $\sigma_{xx}(0)$.

We are interested in the effect on transport of the dynamics of
individual electrons in the electron liquid. Respectively, we will
consider scattering with momentum transfer that largely exceeds the
reciprocal inter-electron distance, $q\gg n^{-1/2}$. Such scattering
is usually more important for magnetoconductivity. If, on the other
hand, the random potential is smooth on the scale of inter-electron
distance,
$r_{\rm c}\gg n^{-1/2}$, the magnetoconductivity can be analyzed in
the magneto-hydrodynamic approximation by considering long-wavelength
hydrodynamic modes of a viscous electron liquid in a magnetic field
and disorder potential.

\subsection{Gaussian potential with
correlation length $r_{\rm c}\ll n^{-1/2}$}

We start with the case of a weak Gaussian random potential with
correlation length small compared to the inter-electron
distance. Here, to the second order in $V({\bf r})$ the conductivity
is given by Eqs.~(\ref{Einstein}), (\ref{tau}), with the structure
factor ${\cal S}({\bf q},\omega)$ evaluated in the single-site
approximation (\ref{eq:single_site}).

The results become particularly interesting and instructive if the
correlation length of the potential satisfies the condition
\begin{equation}
\label{smooth}
\deltaf \ll r_{\rm c}\ll n^{-1/2},
\end{equation}
where $\deltaf$ is thermal electron displacement from a
quasi-equilibrium position in the liquid (\ref{short-range}).

For a random potential that satisfies the inequality (\ref{smooth}),
the structure factor $\tilde{\cal S}_{\rm ss}({\bf q},t)$ needs to be
calculated for $q\ll 1/\deltaf$. In other words, the electron
displacement $|{\bf r}_m(t)-{\bf r}_m(0)|$ in
Eq.~(\ref{eq:single_site}) should exceed $\deltaf$. Such displacements
occur on times $t$ that largely exceed the reciprocal frequency of
vibrations about a quasi-equilibrium position $\Omega^{-1}$. They are
due to self-diffusion in the electron liquid, with the diffusion
coefficient $D_{ee}$ (\ref{eq:diffusion-coefficient-estimated})
introduced in Appendix.  In the diffusion approximation we have
\begin{equation}
\label{S_smooth} \tilde{\cal S}_{\rm ss}({\bf q},t) \approx
\exp(-D_{ee}q^2|t|), \quad |t|\gg \Omega^{-1}.
\end{equation}

The physical picture of scattering by a smooth random potential
$V({\bf r})$ is as follows. The guiding center of the electron
cyclotron orbit drifts transverse to the sum of the many-electron
fluctuational force $-e\ef$ and $-{\bm\nabla}V({\bf r})$. The field
$\ef$ leads primarily to vibrations about a quasi-equilibrium electron
position. In turn, these vibrations result in partial averaging of the
disorder potential. This is somewhat similar to motional narrowing in
nuclear magnetic resonance.
The averaging is incomplete because of self-diffusion of
quasi-equilibrium electron positions. Therefore the momentum
transferred by the disorder potential, and thus the conductivity, are
determined by the diffusion rate.

Frequency dispersion of the conductivity depends on a specific
model of the random potential. A model frequently used in the
analysis of scattering of 2DES, including the quantum Hall effect, is a
random potential with a Gaussian correlator \cite{Huckestein-95},
\begin{equation}
\label{Gaussian}
\overline{|V_{\bf q}|^2}= S^{-1}v^2_G\,\exp(-q^2r_{\rm c}^2/2),
\end{equation}
i.e., $\overline{V({\bf r})V({\bf r}')}=(v_G^2/2\pi
r_{\rm c}^2)\, \exp[-({\bf r-r}')^2/2r_{\rm c}^2]$.

{}From Eqs.~(\ref{Einstein}), (\ref{tau}), (\ref{S_smooth}), and
(\ref{Gaussian}) we obtain for the magnetoconductivity
\begin{eqnarray}
\label{sigma_G}
\sigma_{xx}(\omega)&&= {ne^2v_G^2\over 4\pi k_BT\,m^2\omega_c^2\,r_{\rm
c}^2D_{ee}}\,
\sigma^{}_G(\omega),
\\
\sigma^{}_G(\omega)&&=
1+\tilde\omega\left[\cos(\tilde\omega)\left({\rm Si}(\tilde\omega)-
{\pi\over 2}\right) -\sin(\tilde\omega){\rm Ci}(\tilde\omega\right],\nonumber
\end{eqnarray}
where $\tilde\omega=\omega r_{\rm c}^2/2D_{ee}$, and Si$(z)$ and
Ci$(z)$ are the sine and cosine integral functions, respectively
\cite{Abramowitz}. Eq.~(\ref{sigma_G}) is written for $\omega \geq 0$;
the function $\sigma_{xx}(\omega)$ is even in $\omega$.

\begin{figure}[ht]
\includegraphics[width=3.0in]{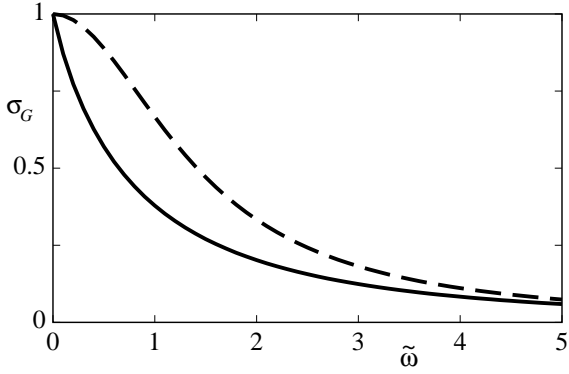}
\caption{Frequency dependence of the reduced microwave many-electron
conductivity $\sigma^{}_G(\omega)$ [Eq.~(\protect\ref{sigma_G})]. The scaled
frequency is $\tilde\omega=\omega r_{\rm c}^2/2D_{ee}$. The dashed
line shows a Lorentzian curve with same maximal value and same
asymptotic behavior for large $\omega$.}
\label{fig:Gaussian}
\end{figure}

The function $\sigma^{}_G(\omega)$ is shown in
Fig.~\ref{fig:Gaussian}. It decays monotonically with the increasing
$\omega$ and has a Lorentzian-type tail, $\sigma^{}_G\approx
2/\tilde\omega^2$ for $\tilde\omega \gg 1$. An unusual feature of the
conductivity is that $\sigma^{}_G$ is {\it linear} in $|\omega|$ for
small $|\omega|$. It has the form $\sigma^{}_G(0)-{\rm const}\times
|\omega|$. This nonanalytic behavior is a consequence of the slow
decay of the structure factor $\tilde{\cal S}_{\rm ss}({\bf q},t)$
(\ref{S_smooth}) with time for small wave numbers $q$. We note that
Eq.~(\ref{sigma_G}) does not apply for very small $|\omega|$. Indeed,
the single-site approximation (\ref{S_smooth}) is valid only for $q\gg
n^{1/2}$. This means that Eq.~(\ref{sigma_G}) can be used only for
$|\omega| \gg D_{ee}n$ or $\tilde\omega \gg nr_{\rm c}^2$.

The typical duration of a collision with the random potential is given
by the time $r_{\rm c}^2/D_{ee}$ it takes for an electron to diffuse
over the length $r_{\rm c}$. The necessary condition for applicability
of the above theory is that this time be small compared to the
relaxation time $\tau(0)$. It sets an upper bound on the strength of
the disorder potential
\begin{equation}
\label{weak_G}
|v_G| \ll \hbar D_{ee}/l.
\end{equation}

The inequality (\ref{weak_G}) can be understood by noticing that, as
mentioned above, a smooth disorder potential adds to the electron
drift velocity a term $i(c/eB)\sum\nolimits_{\bf q}V_{\bf q}(\hat {\bf
z}\times {\bf q})\exp(i{\bf qr})$. Eq.~(\ref{weak_G}) corresponds to
the condition that the root mean square electron displacement due to
this velocity over the collision duration $r_{\rm c}^2/D_{ee}$ be
small compared to the size of the electron wave packet
$l$. Interestingly, the correlation length $r_{\rm c}$ drops out of
Eq.~(\ref{weak_G}).

The parameter range where a 2D electron system forms a liquid is not
limited to temperatures where electron motion is semiclassical,
$k_BT\gg \hbar\Omega$ (\ref{Trange}). The results of this subsection
are valid even for $k_BT < \hbar\Omega$ as long as the electron system
displays self-diffusion. The parameter $\deltaf$ in the inequality
(\ref{smooth}) is in this case determined by quantum zero-point
fluctuations.

The formalism of this subsection can be applied also to the case of a
smooth potential created by charged donors separated from the electron
layer by a spacer of width $d\ll n^{-1/2}$. The major contribution to
the conductivity comes from scattering with momentum transfer $\sim
\hbar/d$. However, the long-wavelength tail of the Coulomb potential
leads to logarithmic divergence of the relaxation rate $\tau^{-1}(0)$
calculated using Eqs.~(\ref{tau}), (\ref{S_smooth}). The analysis of
this divergence requires a hydrodynamic approach and will be carried
out in a separate paper.

\subsection{Electron traps}

An important type of disorder, particularly for MOS systems, are
electron traps. One can think of them as deep short-range potential
wells $V_{\rm tr}({\bf r})$ located at random positions ${\bm
\rho}_{\kappa}$. The overall random potential then is
\begin{equation}
\label{trapping_potential}
V({\bf r}) = \sum_{\kappa} V_{\rm tr}({\bf r}-{\bm\rho}_{\kappa})
\end{equation}
The potentials $V_{\rm tr}$ are not weak, but the trap density $N_{\rm
tr}$ is assumed small. In particular, we assume that the inter-trap
distance $N_{\rm tr}^{-1/2}$ largely exceeds the correlation length of
the electron liquid. Then, even though the traps will capture some
electrons, other electrons will be free to move and there still will
be self-diffusion in the electron liquid.

To lowest order in $N_{\rm tr}$, the magnetoconductivity is given by
Eq.~(\ref{eq:cond_iterated}). For $\hbar\omega\ll k_BT$ we write it
in the form
\begin{eqnarray}
\label{trap_iteration}
&&\sigma_{xx}(\omega)= {e^2l^4\over 4 k_BT \hbar^2 S}
  \,\sum_{n,n'}\sum_{\kappa,\kappa'} \int_{-\infty}^{\infty} dt\,
  e^{i\omega t}\nonumber\\
&&\times \bigl\langle {\bm\nabla}_nV_{\rm
  tr}[{\bf r}_n(t)-{\bm\rho}_{\kappa}]\cdot {\bm\nabla}_{n'}V_{\rm
  tr}[{\bf r}_n'(0)-{\bm\rho}_{\kappa'}] \bigr\rangle.
\end{eqnarray}

Because electrons are strongly correlated, only one electron may
experience the potential of a given trap at a time. We assume that
this electron gets localized on the trap. It then creates a repulsive
Coulomb potential for other electrons, and they stay away from the
trap.

The picture of one localized electron per trap allows us to rewrite the
sum of the correlation functions in Eq.~(\ref{trap_iteration}) as
\begin{equation}
\label{trapping_factor}
\sum_{n,n'}\sum_{\kappa,\kappa'}\langle\ldots\rangle \to
\sum_{\kappa}\langle{\bm\nabla}_{\kappa}V_{\rm tr}\left[{\bf
r}_{\kappa}(t)\right]\cdot {\bm\nabla}_{\kappa}V_{\rm tr}\left[{\bf
r}_{\kappa}(0)\right]\rangle.
\end{equation}
Here, ${\bf r}_{\kappa}$ is the coordinate of an electron on a
$\kappa$th trap counted off from the trap position ${\bm
\rho}_{\kappa}$. We also disregarded terms
with $\kappa'\neq \kappa$, because different traps are far from each
other and electrons on different traps are not correlated.

We will assume that a localized electron occupies only the ground
bound state $|g\rangle$ in the potential $V_{\rm tr}({\bf r})$ and
that the energy spacing between the ground and nearest excited state
is $\hbar\omega_{g\to e}\gg k_BT$. Then electrons very rarely
escape from the traps or are thermally excited to higher states. It should
be noted that, in fact, the true binding potential is stronger than
the ``bare'' potential $V_{\rm tr}$, because surrounding electrons
contribute to electron localization by providing a
``caging'' potential. The localization length of the state $|g\rangle$
in a quantizing magnetic field is of the order of the magnetic length
$l \ll\deltaf$.

For $\hbar\omega_{g\to e}\gg k_BT$, thermal averaging of a
single-electron operator $\hat {\cal O}\equiv {\cal O}({\bf
r}_{\kappa})$ in Eq.~(\ref{trapping_factor}) is done in two
steps. First, one has to find the diagonal matrix element $\langle
g|\hat {\cal O}|g\rangle$ on the wave functions of the ground state of
the trapped electron. Then the matrix element has to be averaged over
the states of the many-electron system. In addition, if we are
interested in the conductivity at frequencies $\omega\ll \omega_{g\to
e}$, we have
%
\begin{eqnarray*}
\langle
g|{\bm\nabla}_{\kappa}V_{\rm tr}\left[{\bf
r}_{\kappa}(t)\right]\cdot{\bm\nabla}_{\kappa} V_{\rm
tr}\left[{\bf r}_{\kappa}(0)\right]|g\rangle\nonumber\\
\to \langle
g|{\bm\nabla}_{\kappa}V_{\rm tr}\left[{\bf
r}_{\kappa}(t)\right]|g\rangle\,\langle g|{\bm\nabla}_{\kappa} V_{\rm
tr}\left[{\bf r}_{\kappa}(0)\right]|g\rangle.
\end{eqnarray*}
%

In order to calculate the matrix elements $\langle
g|{\bm\nabla}_{\kappa}V_{\rm tr}({\bf r}_{\kappa})|g\rangle$ we note
that, besides the potential of the defect, a trapped electron
experiences a potential from other electrons.  The fluctuating part of
this potential varies on time $\sim \Omega^{-1}$. In the range of
interest, $\beta\Omega\ll 1$ (\ref{Trange}) and $\beta\omega_{g\to e}
\gg 1$, we have $\Omega\ll \omega_{g\to e}$. Therefore a localized
electron follows the many-electron field adiabatically. The overall
force on a $\kappa$th localized electron is $-{\bm\nabla}V_{\rm
tr}\left({\bf r}_{\kappa}\right) - e{\bf E}_{\kappa}$, where ${\bf
E}_{\kappa}$ is the many-electron fluctuational field on this electron
(see Appendix~\ref{sec:liquid}).  The diagonal matrix element of the
force should be equal to zero. This gives for the
cor\-re\-lator~(\ref{trapping_factor})
\begin{equation}
\label{long_range_corr}
\langle {\bm\nabla}_{\kappa}V_{\rm tr}\left[{\bf
r}_{\kappa}(t)\right]\cdot{\bm\nabla}_{\kappa} V_{\rm tr}\left[{\bf
r}_{\kappa}(0)\right]\rangle =e^2\langle {\bf E}_{\kappa}(t){\bf
E}_{\kappa}(0)\rangle.
\end{equation}

For a trapped electron, the behavior of $\langle {\bf
E}_{\kappa}(t){\bf E}_{\kappa}(0)\rangle$ is determined by motion of
neighboring electrons. As in the absence of trapping, this motion is
vibrations about quasi-equilibrium positions superimposed on diffusion
of these positions. Because the trapped electron itself does not move,
the vibrations and the diffusion are somewhat different from those in
the free electron liquid. However, we do not expect that this
difference to be too large. Indeed, a neighbor of a trapped electron
has at most one of its six nearest neighbors localized. Therefore the
correlator $\langle {\bf E}_{\kappa}(t){\bf E}_{\kappa}(0)\rangle$
should still decay on times of the order of the reciprocal vibration
frequency $\Omega^{-1}$.

It is seen from Eqs.~(\ref{trap_iteration}), (\ref{long_range_corr})
that $\sigma_{xx}(0)$ is determined by the integral of the field
correlator over time, with no extra time-dependent weight. In the
semiclassical approximation, this integral is simply related to the
self-diffusion coefficient $D_{ee}$ of the electron liquid
(\ref{D_defined}), (\ref{eq:rms-displacement}).  For a trapped
electron, it should be the same, to the order of magnitude, and may
only differ by a factor $\sim 1$ that depends on $\Gamma$. Therefore
\begin{equation}
\label{trapped_conductivity}
\sigma_{xx}(0)\sim N_{\rm tr}e^2D_{ee}/k_BT.
\end{equation}

In contrast to the cases discussed before, the conductivity
$\sigma_{xx}(\omega)$ does not peak at $\omega = 0$ for low
frequencies. The field power spectrum, which is given by the Fourier
transform of the correlator (\ref{long_range_corr}), increases with
$\omega$. A simple calculation shows that, for a Wigner crystal, this
increase is linear for $\omega\ll \Omega$. Therefore for an electron
liquid it should be linear in the range $\deltaf^2/D_{ee}\ll \omega\ll
\Omega$. The conductivity peaks at $\omega\sim \Omega$. The overall
width of the peak of the low-frequency conductivity is $\sim
\Omega$.

Eq.~(\ref{trapped_conductivity}) shows that, for negatively charged
defects, magnetoconductivity is proportional to the defect density,
rather than the electron density $n$. For $\omega=0$, it depends on
$n$ only in terms of the self-diffusion coefficient $D_{ee}$. The
shape of the peak also depends on $n$. Therefore measurements of the
conductivity spectrum should provide an insight into both long- and
short-time electron dynamics in the liquid, i.e., self-diffusion and
vibrations about quasi-equilibrium positions. We note the similarity
between this problem and the problem of dissipative conductivity of a
2D superconducting film with vortices in the presence of pinning
centers.

\section{Conclusions}

In this paper we have found the frequency dependence of the
conductivity of a nondegenerate electron liquid in a quantizing
magnetic field for $\omega\ll \omega_c$. We have shown that this
dependence is extremely sensitive to both short- and long-time
electron dynamics in the liquid and the characteristics of the random
potential.

For a short-range potential, the conductivity is determined by
large-${\bf q}$ electron scattering. It occurs as an electron drifts
transverse to the magnetic field and the field $\ef$, which is created
by density fluctuations in the liquid. The results become particularly
simple if the correlation length of the potential $r_{\rm c}$ is less
than the magnetic length $l$. Here, the shape of the peak of
$\sigma_{xx}(\omega)$ depends on one dimensionless parameter $\omega
t_e$ and is given by an explicit expression
(\ref{sigma_explicit}). The time $t_e$ is the time of flight over the
length $l$ in the field $\ef$. It is smaller than the rate of
inter-electron momentum exchange $\Omega^{-1}$ and depends on the
electron density, temperature, and magnetic field as $t_e\propto
n^{-3/4}T^{-1/2}B^{1/2}$. Therefore by studying the shape of the
magnetoconductivity peak for short-range disorder one can investigate
the short-time electron dynamics as a function of the parameters of
the electron liquid.

The tail of the conductivity in the range $t_e^{-1}\ll \omega\ll
\omega_c$ is exponential and obeys the Urbach rule. The photon energy
$\hbar\omega$ is transferred to the many-electron system via a
radiation-induced electron displacement along the field $\ef$. The
momentum needed for the displacement comes from scattering by a
fluctuation of the disorder potential. As $\omega t_e$ increases, it
becomes more probable for an electron to experience multiple
scattering. This leads to the change of the asymptotic behavior from
$|\ln\sigma_{xx}| \propto \omega$ to $|\ln\sigma_{xx}| \propto
\omega^{2/3}$. Such behavior is described by unusual ``fat fish''
diagrams, which correspond to maximally embedded diagrams in the
self-energy (see Fig.~\ref{fig:fish}).

The conductivity has a different form for a long-range disorder
potential. Of particular interest is the case where the correlation
length $r_{\rm c}$ exceeds the root mean square electron displacement
{}from a quasi-equilibrium position in the liquid $\deltaf$. Here,
$\sigma_{xx}(\omega)$ is determined by electron diffusion in the
liquid. It has a simple form given by Eq.~(\ref{sigma_G}). The shape
of the spectral peak of $\sigma_{xx}(\omega)$ depends on one
dimensionless parameter $\omega r_{\rm c}^2/D_{ee}$. It displays a
smeared cusp at $\omega = 0$ and decays as $\omega^{-2}$ for large
$\omega r_{\rm c}^2/D_{ee}$.

Yet another behavior arises in the case where scatterers are
short-range electron traps. If the density of trapped electrons is
small, the 2DES remains a liquid. The conductivity is expressed in
terms of the power spectrum of the fluctuational field $\ef$ that the
liquid exerts on a trapped electron. The spectral peak of the
conductivity is located at a frequency $\sim \Omega$. The
low-frequency part of the peak is determined by electron diffusion,
whereas the shape of the peak near its maximum depends on vibrations
of electrons about their quasi-equilibrium positions in the liquid.

\begin{acknowledgements} We are grateful to Frank Kuehnel, who
participated in this research at an early stage.  Work at MSU was
supported in part by the NSF through Grants no.  PHY-0071059.
\end{acknowledgements}

\appendix
\section{Dynamics of a nondegenerate 2D electron liquid}
\label{sec:liquid}

A snapshot of a correlated 2D electron liquid is shown schematically
in Fig.~1a. In such a liquid, for most of the time, the electrons
perform small-amplitude vibrations about their quasi-equilibrium
positions in the potential formed by other electrons and the
neutralizing background.  In the absence of a magnetic field, the
characteristic frequency of such vibrations is determined by the
second derivative of the Coulomb potential at the mean inter-electron
distance $n^{-1/2}$ and is given by the short-wavelength plasma
frequency $\omega_p=(2\pi e^2n^{3/2}/m)^{1/2}$, which is the
characteristic Debye frequency of a 2D Wigner crystal.

\begin{figure}[htbp]
    \includegraphics[width=\columnwidth]{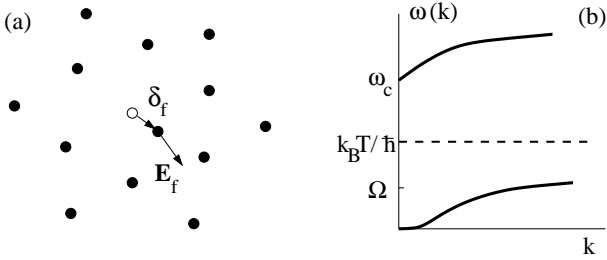}
    \caption{(a) A snapshot of a correlated electron fluid
    (schematically). The open circle shows an equilibrium position of
    one of the electrons in the field of other electrons. Because the
    electron is displaced, it experiences a restoring force, which is
    determined by the fluctuational electric field ${\bf E}_{\rm
    f}$. (b) Phonon spectrum of a 2D Wigner crystal in a quantizing
    magnetic field (schematically).} \label{fig:e-liquid}
\end{figure}

An important characteristic of electron dynamics is the typical
fluctuational electron displacement from the quasi-equilibrium position
$\deltaf$ in Fig.~\ref{fig:e-liquid}a. If the electron motion is
classical, i.e., for $\hbar\omega_p\ll k_BT$, it can be estimated in
the harmonic approximation by setting the potential energy
$e^2n^{3/2}\deltaf^2$ equal to $k_BT$,
\begin{equation}
\label{delta}
\langle\deltaf^2\rangle = k_BT/e^2n^{3/2}.
\end{equation}
The necessary condition that $\deltaf^2$ be much less than the
squared inter-electron distance $n^{-1}$ is equivalent to $\Gamma\gg
1$.

The restoring force on the vibrating electron is given by the electric
field $\ef$, see Fig.~\ref{fig:e-liquid}a. This field is due to
electron density fluctuations. In the classical regime, these are
primarily short-wavelength fluctuations \cite{Dykman-All}. For large
$\Gamma$ the field should be close to its value estimated in the
harmonic approximation in electron displacements from quasi-equilibrium
positions,
%
\begin{equation}
\label{fluct_field}
\langle \ef^2\rangle = F(\Gamma)n^{3/2}k_BT.
\end{equation}
The function $F(\Gamma)$ was obtained by Monte Carlo simulations
\cite{Fang-Yen-97}. In the whole range $10 < \Gamma < \Gamma_W=130$ it
remains essentially constant, varying from 10.5 to 9.1. The field
distribution $p(\ef)$ is Gaussian in the central part,
\begin{equation}
\label{field_distribution}
p(\ef) = [\pi\langle \ef^2\rangle]^{-1}\exp\left[-\ef^2/\langle
\ef^2\rangle\right],
\end{equation}
which is an indication that, in this range of $\Gamma$, electron
motion is mostly weakly anharmonic vibrations. On the far tail, the
decay of $p(\ef)$ is slowed down compared to
Eq.~(\ref{field_distribution})  \cite{Fang-Yen-97}. The
fluctuational field is the only characteristic of the electron liquid,
which is needed in order to describe the magneto-conductivity
$\sigma_{xx}(\omega)$ in the case of weak short-range disorder.

\subsection{A nondegenerate electron liquid in a strong magnetic field}

In a strong magnetic field, $\omega_c\gg \omega_p$, the electron
motion is separated into cyclotron motion with frequencies
$\sim\omega_c$ and comparatively slow vibrations of the guiding
centers about their slowly diffusing quasi-equilibrium positions. The
coordinates of the guiding centers are
\begin{equation}
\label{R_defined}
{\bf R}_n=(X_n,Y_n), \quad {\bf R}_n = {\bf r}_n+
\hbar^{-1}l^2{\bf p}\times\hat{\bf z},
\end{equation}
where,
$\hat{\bf z}=-{\bf B}/B$ is the unit vector normal to the electron
layer, and ${\bf p}$ is the canonical momentum.  From
(\ref{R_defined}), the components $X_n,Y_n$ obey the commutation
relation
\begin{equation}
[X_{n}, Y_{n}]=-i l^2,\quad l^2=\hbar /(m\omega_c).
\label{eq:commutation-relation}
\end{equation}
The magnetic length $l$ gives the typical size of the electron wave
packet.

The dynamics of the guiding centers is described by the Hamiltonian of
the electron-electron interaction $H_{ee}$ (\ref{H_ee}) projected on
the lowest Landau level,
\begin{eqnarray}
  \label{eq:coulomb-hamiltonian}
  H_{ee}\approx {e^2\over2}\sum_{n\neq m}|{\bf R}_n-{\bf R}_m|^{-1}.
\end{eqnarray}

Heisenberg equations of motion for ${\bf R}_n$ can be written in a
closed form in the important case where the electric field on
electrons is smooth on the scale $l$,
\begin{equation}
\label{drift}
\dot{\bf R}_n = cB^{-2}{\bf E}_n\times{\bf B}.
\end{equation}
Here,
\begin{equation}
  \label{eq:field}
  {\bf E}_n = {e}\,{\partial \over \partial {\bf R}_n}
  \sum_{m<n} |{\bf R}_n-{\bf R}_m|^{-1}.
\end{equation}
is the field on the $n$th electron created by other electrons
and calculated ignoring
noncommutativity~(\ref{eq:commutation-relation}) of the guiding
centers' components.

By linearizing ${\bf E}_n$ in displacements of the electrons from
their quasi-equilibrium positions, one can see that the motion of the
guiding centers is mostly vibrations
with typical frequency
\begin{equation}
  \label{Omega}
  \Omega=\omega_p^2/\omega_c = 2\pi e^2n^{3/2}/m\omega_c.
\end{equation}
This frequency also gives the typical rate of inter-electron momentum
exchange.  In the case where electrons form a Wigner crystal, $\Omega$
is the zone-boundary frequency of the lower phonon branch, see
Fig.~\ref{fig:e-liquid}b.

Motion of the guiding centers ${\bf R}_n$ becomes semiclassical for
\begin{equation}
\label{classical}
k_BT\gg
\hbar\Omega.
\end{equation}
It is determined by thermal fluctuations. From (\ref{classical}), a
typical electron displacement from a quasi-equilibrium position
$\deltaf$ given by Eq.~(\ref{delta}) is $\deltaf \gg l$. Therefore the
fluctuational electric field $\ef$, which varies on the distance
$\deltaf$, is uniform on the magnetic length $l$, as assumed in
Eq.~(\ref{drift}).

The probability distribution of the guiding centers $\rho({\bf
R}_1,\ldots, {\bf R}_n,\ldots)$ is given by the Boltzmann equation,
\begin{equation}
\label{Boltzmann}
\rho({\bf R}_1,\ldots,
{\bf R}_n,\ldots) = {\rm const}\times \exp(-H_{ee}/k_BT),
\end{equation}
with $H_{ee}$ given by Eq.~(\ref{eq:coulomb-hamiltonian}). Therefore
the results for a classical electron liquid in the absence of a
magnetic field can now be carried over to the case of quantizing
magnetic field. In particular, the instantaneous distribution of the
fluctuational field $\ef$ is the same as in the classical 2DES for
$B=0$. The condition (\ref{classical}) is much less restrictive than
$\hbar\omega_p \ll k_BT$, because $\Omega\ll \omega_p$.

For $\Gamma < 130$, Eqs.~(\ref{drift}), (\ref{eq:field}),
(\ref{Boltzmann}) provide a semiclassical description of a
nondegenerate electron liquid in a strong magnetic field. In the
long-wavelength limit, this liquid can be alternatively described in
the hydrodynamic approximation. The transport coefficients (e.g.,
viscosity) can be found from the correlation functions of the liquid
(e.g., current-current correlator). In two dimensions, because of a
large contribution from long-wavelength modes, the transport
coefficients diverge. A self-consistent analysis in the case of a
classical 2D liquid for small frequencies $\omega$ and wave numbers
$q$ was done by Andreev \cite{Andreev80}. He showed that transport
coefficients diverge as $\ln^{1/2} \omega$ in the limit of small
$\omega$. For a 2D electron liquid in a short-wavelength random
potential this divergence is terminated at the cutoff frequency given
by the rate of electron scattering in this potential $\tau^{-1}$. This
rate determines the long-wavelength static conductivity and thus the
decay rate of long-wavelength modes of the electron liquid. A
corresponding self-consistent analysis in the presence of a magnetic
field will be given elsewhere. A logarithmic correction to the
viscosity of a 2D Fermi liquid due to electron-impurity scattering was
found in Ref.~\onlinecite{Spivak02}.

An important feature of the liquid state, which ultimately gives rise
to a nonzero static conductivity, is self-diffusion.  The coefficient
of self-diffusion $D_{ee}$ can be related to the r.m.s.\
dis\-place\-ment of a particle over a long time  $t\gg\Omega^{-1}$,
\begin{equation}
\label{D_defined}
D_{ee}={ \Delta R_n^2(t)/4t}, \quad
\Delta R_n^2(t)\equiv\langle [{\bf R}_n(t)-{\bf
    R}_n(0)]^2\rangle.
\end{equation}
If we assume that the semiclassical approximation (\ref{drift})
applies for long times $t\gg \Omega^{-1}$, then the electron
displacement is related to the correlator of the fluctuational
electric field by
\begin{equation}
  \Delta R_n^2(t)
  =(c/B)^2\int_0^t dt'\int_0^t dt''\, \langle{\bf E}_n(t'){\bf
    E}_n(t'')\rangle.
  \label{eq:rms-displacement}
\end{equation}
The natural scale of the electric field is given by the
r.m.s.~value~(\ref{fluct_field}). Field correlations decay over the
time $\sim \Omega^{-1}$.  Then, the diffusion coefficient becomes
\begin{equation}
  \label{eq:diffusion-coefficient-estimated}
  D_{ee}={k_B T\over m\omega_c}\,\tilde D_{ee}(\Gamma, \Omega t).
\end{equation}
Here, we took into account that $c^2\langle \ef^2\rangle/B^2\Omega
\sim k_BT/m\omega_c$.

We note that, in the absence of a magnetic field, the power spectrum
of the field $\ef$ goes to zero for $\omega\to 0$.  In this case, the
double integral over time in Eq.~(\ref{eq:rms-displacement}) gives
just the average increment of the momentum of an electron, which
saturates for large $t$. The electron dynamics in a magnetic field is
different, and we expect that here the integral
(\ref{eq:rms-displacement}) linearly increases with $t$. This
conjecture is based on the argument that, because of self-diffusion in
the electron liquid, $\Delta R_n^2(t)$ should linearly increase in
time. This increase must be caused by the fluctuational field, since
quantum corrections to an electron displacement are small
\cite{Dykman-All}.

An analysis based on magneto-hydrodynamics with a
frequency-independent viscosity coefficient leads to an extra factor
$\ln t$ in the time dependence of $\Delta R_n^2(t)$. A similar factor
should arise in the function $\tilde D_{ee}$ for $t \ll \tau$. However,
it should saturate and become a constant for $t> \tau$.  For a
correlated liquid, we expect that the factor $\tilde D_{ee}$ is not
large,
$\tilde D_{ee}\lesssim 1$.

In the absence of a magnetic field, straightforward scaling arguments
give for the self-diffusion coefficient an expression of the type
(\ref{eq:diffusion-coefficient-estimated}), with $\omega_c$ replaced
by the characteristic vibration frequency $\omega_p$. The numerical
factor which stands for $\tilde D_{ee}$ in this expression is known
{}from the data of
simulations~\cite{Hansen79,Kalia81,Strandburg88,Fang-Yen-97}, it is
$\sim 0.1$ close to the melting transition and increases with
temperature.

\end{document}